\documentstyle[prb,aps]{revtex}
\begin{document}
\def\baselinestretch{1.3}
\draft
\large
\title{\Large\bf Current-Voltage Characteristics of Polymer
Light-Emitting Diodes}
\bigskip

\author{\large E. M. Conwell$^{a,b}$ and M.W. Wu$^a$}
\address{\large \mbox{}$^a$ Center for Photoinduced Charge Transfer,
Chemistry Department,
University of Rochester, Rochester, New York 14627\\
\mbox{}$^b$ Xerox Corporation, Wilson Center for Technology, 114-22D,
Webster, New York 14580}
\maketitle
\bigskip
\bigskip

\centerline{\large\bf ABSTRACT}
\smallskip

 Conduction in pristine conjugated polymers (other than polyacetylene)
is by polaron hopping between sites corresponding to
conjugation lengths. The strong increase of current $I$ with voltage $V$
observed for both emission-limited and ohmic contacts is due in large
part to mobility increase as increasing field makes it more possible
to overcome internal barriers, such as energy differences
between sites. For emission-limited contacts an additional source
of nonlinear increase of $I$ with increasing $V$ is greater ability to
escpe return to the injecting electrode due to the image force.
For ohmic contacts additional nonlinearity comes from space charge
effects. We are able to fit $I$ {\em vs.} $V$ for electron or
hole conduction in some poly($p$-phenylene vinylene), PPV,
derivatives with ohmic contacts for
reasonable values of the parameters involved.
\bigskip

\bigskip

\centerline{\large\bf 2. INTRODUCTION}
\smallskip

Contact injection into conducting polymers has usually been described
as tunneling of electrons or holes into a broad conduction or
valence band, respectively, in the polymer through interface
barriers arising from the band offset between the polymer and the
metal electrodes.\cite{1,2} This picture must be corrected in
several respects. First, the existence of broad continuous bands requires
essentially infinite conjugation length whereas, it is
generally acknowledged, typical conjugation lengths of PPV, for
example, are $\sim$ 6 or 7 monomers. For the typical conjugation
length the average spacing between levels is $\sim$ several $kT$ at
room temperature.\cite{3,4} A more essential objection is
that the carriers are much more likely to tunnel into polaron levels
than into the discrete levels in the ``conduction  band'' or
``valence band'' because the polaron levels are lower in energy.
The polaron levels available for electron injection lie below the
lowest unoccupied molecular orbital (LUMO), while those available for
hole injection lie above the highest occupied molecular
orbital(HOMO). The separation of the polaron level from the
LUMO or HOMO depends on chain length. For PPV, calculations give the
separation as 0.15\ eV,\cite{5} or 0.2\ eV,\cite{3} for very long
segments, increasing to $\sim$ 0.7\ eV for a 3-monomer long segment.\cite{3}
Although the polaron levels do not exist unless they are occupied by  an
electron or hole, tunneling into them may be facilitated by preparation of
the appropriate chain deformation through the large zero point fluctuations
characteristic of quasi-one-dimensional materials.

Once an electron or hole tunnels into a polaron level near the contact, under
the influence of the electric field it moves toward the other contact,
hopping from one conjugation length to another. The variation in
conjugation length and the presence of defects result in a
spread in energy of the hopping sites (diagonal disorder).
Additional disorder is provided by the random
orientation and variations in separation
of the sites (off-diagonal disorder).
As has been pointed out by a number
of authors, the situation for
hopping transport in a
conjugated polymer
is quite similar
to that in
molecularly doped
polymers. Strong evidence
for the similarity comes from
the fact that the field dependence
of the mobility $\mu$ generally observed
for molecularly doped polymers,\cite{6}
\begin{equation}
\label{eq1}
\mu=\mu_0 e^{\alpha E^{1/2}}\;,
\end{equation}
$\alpha$ being a constant, has now been observed
for PPV and a couple of its derivatives.\cite{7,8,9}
This field dependence is a result of disorder. An appropriate
model for treating transport in these systems for low injection
is the disorder model pioneered by B\"assler and associates.\cite{10}
The distribution in energy of hopping sites, in this case the polaron states,
is taken as a Gaussian with variance $\sigma$.\cite{10} Comparison of disorder
theory with various experimental data for PPV leads to $\sigma\sim 0.1$\ eV
for that case.\cite{11}

The behavior of a contact depends critically on the location of the
polaron states relative to the Fermi energy, $E_F$, of the metal. Internal
photoemission measurements, such as those of Campbell {\em et al.},\cite{12}
yielding the energy required to inject an electron from a metal into a
polymer, give the energy difference between $E_F$ and some average
state in the polaron distribution. We denote this energy by $W$. In section II
we discuss $I$-$V$ characteristics for a case of large $W$, which means small
injection, using the results of a Monte Carlo simulation based on the disorder
model.\cite{13} In section III we carry out calculations for $W\sim0$, or
ohmic contacts, first for the case of only one contact injecting,\cite{14}
then for both contacts injecting. We use the classical approach of Lampert for
both cases.\cite{15,16} In the former case we obtain good agreement with
the current $I$ {\em vs.} voltage $V$ characteristic of
poly(2-methoxy, 5-(2$^\prime$-ethyl-hexyloxy)-1, 4 $p$-phenylene
vinylene], MEH-PPV, samples with only electrons injected, from a Ca
contact\cite{12} for which photoemission gives $W\sim0$.\cite{12}
This calculation requires taking into account trapping of the electrons.
A calculation of $I$ {\em vs.} $V$ for electrons only injected, including
trapping, was carried out for poly(dialkoxy phenylene vinylene).\cite{17}
The steep experimental  variation of $I$ with $V$ was fitted by assuming
a distribution of traps exponential in energy. These results are not
meaningful, however, because mobility was taken as constant rather than
varying with field according to Eq.\ (\ref{eq1}). In later work by some of
the same authors calculations were carried out for poly(dialkoxy PPV) samples
with hole injection only, for which trapping may be neglected, with $\mu$
given by Eq.\ (\ref{eq1}) and the contacts assumed ohmic.\cite{9}
These calculation give good agreement with experimental $I$ {\em vs.} $V$
despite the fact that $W\not=0$ for the ITO contact that was used. The
agreement supports the assumption that this contact was nevertheless not
emission-limited for the range of fields in the study.\cite{9}
\bigskip

\centerline{\large\bf 3. $I$-$V$ CHARACTERISTIC FOR LARGE $W$}
\smallskip

Large $W$ implies emission-limited contacts, which in turn usually means
space charge effects may be neglected. Current variation with field may
then be obtained by the usual Monte Carlo simulation for transport in
disordered materials. We followed the treatment of Ref.\ 13. Taking
$E_F$ of the metal as the zero of the energy scale and the metal-polymer
interface as $x=0$, we may write the energy $U$ of a polaron site in the
electric field:
\begin{equation}
\label{eq2}
U(x)=W-eEx-e^2/4\kappa x
\end{equation}
where the last term represents the image force, $\kappa$ being
the dielectric constant, taken as 3. In the presence of energetic disorder
$U(x)$ gives the mean energy $\bar\varepsilon(x)$ of a polaron
site at $x$. The energy $\varepsilon$ is assumed to have a
Gaussian-distribution around $\bar\varepsilon$ given by
$P(\epsilon,\bar\epsilon)=(2\pi\sigma^2)^{-1/2}\exp[-(\varepsilon
-\bar\varepsilon)^2/2\sigma^2]$. The rate of carrier tunneling into a
polaron state of energy $\varepsilon$ at $x$ is assumed to be
$v_m(x)e^{-\varepsilon/kT}$, where the last factor represents
the probability of the carrier having thermal energy $\varepsilon$
above $E_F$. The total number of carriers tunneling into polaron levels
at $x$ per second is then $v_m(x)\exp[-(\bar\varepsilon(x)-\sigma^2/
2kT)/kT]$.\cite{13} It is reasonable to assume that this represents the
initial population in the first layer of the polymer. Thus, as expected for
large $W$, the mean energy of the initially populated sites is
below $\bar\varepsilon$, specifically by the amount $\sigma^2/kT$. It is
apparent from this that increase in $\sigma$, {\em i.e.}, broadening
the energy distribution of transport sites by disorder, increases
the current tunneling into the polymer. This effect was found experimentally
by Vestweber {\em et al.},\cite{18} who noted that the LED current
increased with increase in $\sigma$, indicating that the increased
injection dominates over the decreased $\mu$ that also results from increase
in disorder.

Given the rate of carriers initially tunneling into the polymer
and their distribution in energy, we can determine by Monte Carlo simulation
the yield, {\em i.e.}, the fraction of these carriers that escapes the
return to the electrode and reaches the other boundary of the sample. Even
in the absence of disorder the image force results
in the great majority of carriers returning to the electrode at low fields.
An analytic solution for small injection (space charge neglected) in the
ordered case based on Eq.\ (2),\cite{19,20} gives the yield as 0.3 \% in
a field of 1.25$\times 10^5$\ V/cm at 300\ K for the parameters of Ref. 13.
The results for the Monte Carlo simulation for $\sigma=0$ were in good
agreement with the analytic solution.\cite{13} In the disordered
case the carriers have to overcome the random barriers due to disorder as
well as the image force. The result is that at fields $\sim 10^5$\ V/cm
mild disorder ($\sigma=0.08$\ eV)  results in a yield smaller by a couple of
orders of magnitude\cite{13} even though as pointed out above, large $\sigma$
makes the injection larger. In addition the yield increases more strongly
with increasing field, at least partially the result of increasing $\mu$
with increasing field.

To calculate current {\em vs.} field it is necessary to multiply the
yield by the number of carriers entering the polymer per second. The
result of this for the parameters of Ref. 13 and a sample length of 120 nm
is a current increase by a factor of 10$^5$ as the applied voltage goes from 1
to 20 eV.\cite{14} For PPV the increase obtained should be even larger because
indications are that $\sigma\sim0.1$ rather than 0.08\ eV.
\bigskip

\begin{center}
{\large\bf 4. $I$-$V$ CHARACTERISTIC FOR OHMIC INJECTING  CONTACTS}
\end{center}

We consider first the case where only electrons are injected. Possion's
equation may then be written
\begin{equation}
\label{de}
(\kappa/e)dE/dx=\rho
\end{equation}
where
\begin{equation}
\rho=(n-\bar n)+(n_t-\bar n_t)\;.
\end{equation}
Here $e$ is the charge on the electron, $n$ and $n_t$ the densities
of free and trapped electrons, respectively, and $\bar n$ and
$\bar n_t$ their respective average values for the sample in
thermal and electrical equilibrium (no applied voltage). Following
Lampert\cite{15} we simplify the equation for current density
$J$ by neglecting the diffusion terms. For electrons only,
with Eq.\ (1) incorporated, the steady current density is given by
\begin{equation}
\label{j}
J=ne\mu_0\exp(\alpha E^{1/2})E\;.
\end{equation}
With the simplification of neglecting the diffusion current
the boundary condition at the cathode interface $x=0$ is $E=0$.\cite{15}
$I$-$V$ characteristics with only electrons injected were obtained
by Parker on MEH-PPV samples with one Ca electrode and
the other electrode Nd, Mg or Ca. As shown in Ref. 14 we were able
to fit Parker's data with the calculations described above for both
trap-free ($n_t=\bar n_t=0$) and all-traps-filled ($n_t=N_t$, the total
trap density) cases.\cite{14} In the latter case there is a threshold for
current flow because the repulsion of the charges in the traps prevents
injection of additional charge at low fields. However, beyond the
threshold, above $\sim 5$ V or 4$\times 10^5$\ V/cm for the particular
samples, there is a good fit to the data under the assumption that all
traps are filled. For the trap-free case the parameters for the
fit were $\mu_0=5\times 10^{-11}$ cm$^2$/Vs, $\alpha=8\times 10^{-3}$
cm$^{1/2}$/V$^{1/2}$. For the trap-filled case the best fit was
obtained with $\mu_0=5\times 10^{-9}$\ cm$^2$/Vs, $\alpha=4.5
\times 10^{-3}$ cm$^{1/2}$/V$^{1/2}$ and $N_t=10^{17}$/cm$^3$. (It is
reasonable to assume $\bar n=\bar n_t=0$). The latter set of parameters,
particularly $\mu_0$, appears to be in better agreement with values
determined by other methods. Hole mobility {\em vs.} field, measured
for PPV by observing the time delay after injection for the appearance
of luminescence, yields by extrapolation $\mu_0=5\times 10^{-9}$
cm$^2$/Vs.\cite{7} Electron mobility  in PPV and its derivatives is
generally found at not too high fields to be a couple of orders of
magnititude smaller than hole mobility, $\mu_p$, the differences
attributed to trapping, by carbonyls for example.\cite{11}
It is reasonable that $\mu_n$ increases as $E$ increases and thus
trap-filling increases. The finding that $\mu_n\sim\mu_p$ in the
trap-filled limit is not unexpected because bandstructure is similar
for electrons and holes in PPV.\cite{21} It is interesting to note that
$\mu_0$ is different for different PPV derivatives. For dialkoxy-PPV,
for example, Blom {\em et al} obtained $\mu_0=5\times 10^{-7}$\ cm$^2$/Vs
for holes.\cite{17} We suggest that the differences reflect different degrees
of disorder. Support for this is the fact that in dialkoxy-PPV
$\mu$ can be measured by tranditional time-of-flight technique, in which
the arrival time at the other electrode of an injected pulse of
carriers is observed, whereas transport in PPV and MEH-PPV is too
dispersive for that method to work. The value of $\alpha$ found by
Karg {\em et al.} is $6\times 10^{-3}$\ cm$^{1/2}$/V$^{1/2}$,\cite{7}
close to the value of $4.5\times 10^{-3}$\ cm$^{1/2}$/V$^{1/2}$
found here, and also to the value $5.4\times 10^{-3}$\ cm$^{1/2}$/V$^{1/2}$
found for dialkoxy-PPV.\cite{9} The value of $N_t$ for the particular
sample used is, of course, unknown but is similar to the value
estimated by Campbell {\em et al.} for MEH-PPV from the change in
capacitance with forward bias.\cite{22} From the good fit at high voltage we
conclude that the current there is space-charge limited. It does not
vary as $V^2$ because of the strong field-dependence of $\mu$.

We consider now the case of both electrons and holes injected. For
this analysis we simplify Possion's equation by taking
\begin{equation}
\label{rho}
\rho=n-p+N_d\;,
\end{equation}
where $N_d$ is the density of traps filled by electrons. For the current
density we use Eq.\ (\ref{j}) with $\mu_0$ replaced by $\mu_n^0$,
and a similar term added for holes with $\mu_n^0$ replaced by $\mu_p^0$.
 We assume $\alpha$ is the same for electrons and holes. Steady state
requires that
\begin{equation}
\frac{dJ}{dx}=\frac{dj_n}{dx}+\frac{dj_p}{dx}=0
\end{equation}
where $j_n=n\mu_n(E)E$ and $j_p=p\mu_p(E)E$. For the steady state, on the
simplifying assumption that the lifetime for recombination $\tau$ is a
constant, independent of $x$, we obtain\cite{16}
\begin{eqnarray}
\label{djn}
dj_n/dx&=&-n(x)/\tau_n\;,\\
\label{djp}
dj_p/dx&=&n(x)/\tau_n\;.
\end{eqnarray}
Eliminating $n$ and $p$ from Eqs.\ (\ref{de}), (\ref{rho}),
(\ref{djn}) and (\ref{djp}), and using Eq.\ (\ref{eq1}), we
obtain
\begin{equation}
Ee^{\alpha\sqrt{E}}\frac{d}{dx}(Ee^{\alpha\sqrt{E}}\frac{dE}{dx})
-\frac{1}{\mu_p^0\tau_n}Ee^{\alpha\sqrt{E}}\frac{dE}{dx}
=-\frac{J}{\epsilon\mu_p^0\mu_n^0\tau_n}\;.
\end{equation}
This equation can be solved by assuming
\begin{equation}
\label{tr}
Ee^{\alpha\sqrt{E}}\frac{d}{dx}=\frac{d}{dy}\;.
\end{equation}
The solution is
\begin{equation}
E(y)=C_1e^{by}-(A/b)y+C_2
\end{equation}
where $b=1/\mu_p^0\tau_n$, $A=-J/\epsilon\mu_p^0\mu_n^0\tau_n$
and $C_1$, $C_2$ arbitrary constants to be determined by the
boundary conditions.

We assume both contacts are ohmic, so the condition at the cathode is
\begin{equation}
E(y=0)=E(x=0)=0
\end{equation}
and at the anode
\begin{equation}
E(x=L)=E(y=y_c)=0
\end{equation}
where $y_c$ is determined from
\begin{equation}
L=\int_0^{y_c}E(y)e^{\alpha\sqrt{E(y)}}dy\;.
\end{equation}
With these boundary conditions
\begin{equation}
C_1=(A/b)y_c(e^{by_c}-1)^{-1},\hspace{1cm}C_2=-C_1\;.
\end{equation}
The applied voltage is given by
\begin{equation}
V=\int_0^LE(x)dx=\int_0^{y_c}E^2(y)e^{\alpha\sqrt{E(y)}}dy\;.
\end{equation}

These equations have been used to fit data for an MEH-PPV diode
with one Ca contact and the other contact Au or ITO, all of
which may reasonably be considered ohmic. Shown in Fig. 1 is the fit for
$N_d=0$ and the parameters for that fit. It should be noted that
the $\mu_n^0$ value is similar to that obtained earlier for the
trap-free single carrier case. To further elucidate the meaning of
the results we calculated $j_n$ and $j_p$, the electric field, $n$, $p$
and the ratio $p/n$ as functions of $x$ for a particular current density,
$J=0.2$\ mA/mm$^2$ corresponding to $V=7.4$\ V. It was found that
the large ratio $\mu_p^0/\mu_n^0$ causes $p>>n$ over 2/3 of the sample.
This results in $j_p>>j_n$ except in the immediate vicinity of the
cathode. To maintain constant current given the large ratio of $\mu_p$
to $\mu_n$ requires that the electric field be larger near the cathode than
the anode, rising more rapidly from $E=0$ at the cathode than at the
anode. If we allow $N_d\not=0$ and $\mu_p=\mu_n$ as
expected for the trap-filled case, the curves become more symmetrical in
$n$ and $p$ and the field is more symmetrical about the middle of the
sample.
\bigskip

\centerline{\large\bf 5. CONCLUSIONS}
\smallskip

Because of the short conjugation lengths, characteristic of PPV and other
conducting polymers, there are no broad continuous bands and
carriers are most  likely to tunnel into polaron levels with little or
no hindrance by a barrier. Random variation in conjugation lengths, and
thus site energies, as well as other sources of disorder, result in the
hopping mobility of polarons increasing exponentially with $\sqrt{E}$,
as has been shown exponentially for several PPV derivatives. This increase of
$\mu$ with $E$ contributes to the strong increase of $I$ with $V$ observed
for LEDs, for example. We have studied the variation of $I$ with $V$ for
polymer samples with emission-limited contacts and with ohmic contacts. In
the former case we did not assume any $\mu$ variation with $E$ but relied
on the usual Monte Carlo simulation to determine first the number
of carriers that make it across the sample to the other electrode as
a function of $E$, and then the current. We find that $I$ increases steeply as
$V$ increases, both because carriers can better overcome the image force
and more easily surmount barriers
due to random site energy variations.

In the case of ohmic contacts, space charge effects also contribute to the
steepness of $I$ {\em vs.} $V$. It is now documented that the $I$ {\em
vs.} $V$ characteristic is well fitted by the use of Poisson's Equation and
the current continuity condition, with insertion of the $\mu$
dependence on $E$, for hole only transport in dialkoxy PPV.\cite{9}
Fits to Parker's $I$ {\em vs.} $V$ data for hole transport only
and electron transport only in MEH-PPV,\cite{1} suggest that trapping
is important in both those cases. A good fit has
been obtained for high enough fields that all traps should be filled, with
parameters that are in reasonable agreement with the
values found from other experiments. More work is needed to fit
$I$ {\em vs.} $V$ for the case where both carriers contribute.
\bigskip

\centerline{\large\bf ACKNOWLEDGEMENTS}
\smallskip

We acknowledge the support
of the National Science Foundation
under Science and Technology Center grant CHE912001.

\def\baselinestretch{1.1}

\newpage
\mbox{}
\vskip4.7in

\noindent Fig.1. Fit of theory (solid curve) to data (dots) of Ref. 1 (Fig. 13)
for MEH-PPV with electrons injected at Ca contact, holes injected
from ITO. The parameters for the fit are $N_d=0$, $\alpha=7\times 10^{-3}$
cm$^{1/2}$/V$^{1/2}$, $\mu_p^0=10^{-9}$ cm$^2$/Vs, $\mu_n^0=0.01\mu_p^0$,
$\tau=1.4\times 10^{-4}$ s.

\end{document}